# Magnetic-Field-Induced Change of Magneto-Electric Coupling in the in Molecular Multiferroic $(ND_4)_2[FeCl_5(D_2O)]$

J. Alberto Rodríguez-Velamazán,[1,2] Oscar Fabelo,[1*] Javier Campo,[2] Angel Millan,[2] Juan Rodríguez-Carvajal,[1] and. Laurent C. Chapon[1*]

[1] *Institut Laue-Langevin, 71 Avenue des Martyrs, CS 20156, 38042 Grenoble Cedex 9, France.*

[2] *Instituto de Ciencia de Materiales de Aragón, CSIC-Universidad de Zaragoza, C/ Pedro Cerbuna 12, E-50009, Zaragoza, Spain.*

*Corresponding Authors:* fabelo@ill.fr *and* chapon@ill.fr

The possibility of cross-control of electric order by magnetic fields and magnetic order by electric fields is the key of the noticeable interest attracted by magneto-electric (ME) multiferroics, since it represents both a great potential for applications and a complex and fascinating topic for fundamental studies [1,2,3,4,5,6,7]. Two types of multiferroic materials are generally distinguished: "type I", where electric and magnetic orders coexist but are weakly coupled, and "type II", where electric polarization appears as a consequence of the magnetic ordering. In this last class, the competition between different magnetic exchange couplings often yields a complex ordered state that breaks inversion symmetry and induces ferroelectric polarization. This coupling scheme is realized through different mechanisms, which can be grouped into three major models: [8] (i) exchange striction, (ii) spin current, and (iii) spin-dependent p-d hybridization (which does not involve an exchange coupling but a single-ion term). For the exchange striction mechanism (i) to take place, a striction along a specific crystallographic axis may be induced by the symmetric exchange interaction between two neighboring spins, $S_i$ and $S_j$, which couples to the pre-existing dipoles. Electric polarization, **P**, will be macroscopically observable if the induced striction is not cancelled after the sum over the bonds of the crystal lattice. This is the mechanism responsible for the induced polarization in compounds with collinear up-up-down-down spin structure ('E-type' antiferromagnetic structure)[9]. These compounds display the highest values of spin-driven ferroelectric polarization [10]. The spin current model (ii) is, in turn, probably the most thoroughly studied. This mechanism is responsible for the magneto-electric coupling in compounds with cycloidal spin arrangements such as $RMnO_3$ (R= rare earth) [11,12,13] $MnWO_4$ [14,15] or $CoCr_2O_4$ [16]. According to this model, the antisymmetric Dzyaloshinskii-Moriya interactions [17,18] produce an electric polarization, **P**, proportional to $\mathbf{r}_{ij} \times (\mathbf{S}_i \times \mathbf{S}_j)$, where $\mathbf{r}_{ij}$ is the vector connecting the nearest spins, $\mathbf{S}_i$ and $\mathbf{S}_j$, and $(\mathbf{S}_i \times \mathbf{S}_j)$ is the so-called spin-chirality vector [19, 20]. Finally, the spin-dependent *p-d* hybridization model (iii) relies on the hybridization of the *d* orbitals of a magnetic site *i* with the *p* orbitals of a neighboring ligand *l*, due to the spin-orbit coupling. The covalence between orbitals is modulated depending on the spin direction, and thus a local **P**, proportional to $(\mathbf{S}_i \cdot \mathbf{e}_{il})^2 \mathbf{e}_{il}$, is induced along the bond direction $\mathbf{e}_{il}$, giving rise to a macroscopic **P** for specific symmetries allowing contributions on each site to add-up. This mechanism is at the origin of electric polarization, for example, in antiferromagnets with proper screw spin arrangement (where the spin current mechanism does not predict electric polarization) like $CuFeO_2$ [21] or $RbFe(MoO_4)_2$ [22], where the magnetic structure is coupled to structural axiality (ferroaxial coupling) [23].

In this letter, we elucidate the changes of magneto-electric coupling mechanism in different zones of the rich magnetic field-temperature (*B-T*) phase diagram of the molecular multiferroic $(NH_4)_2[FeCl_5(H_2O)]$[24], which represents one of the rare cases where improper ferroelectricity has been observed in a molecular material. The multiferroicity of this compound [25] was recently described in a thorough study of its macroscopic physical properties [24]. Later on, we determined the mechanism of multiferroicity in zero magnetic field in the deuterated form of this material from a detailed determination of its crystal and magnetic structures by neutron diffraction [26]. The proposed magnetic structure in zero magnetic field and 2 K corresponds to a cycloidal spin arrangement propagating parallel to the *c*-axis and with magnetic moments mainly contained in the *ac*-plane, thus producing a ferroelectric polarization, primarily directed along the *a*-axis, through the spin current mechanism induced via the inverse Dzyaloshinskii-Moriya interaction [26]. Whilst this system shows a sequence of magnetic transitions very similar to that found in $TbMnO_3$, the archetype of cycloidal multiferroics, it appears that the change of electric polarization under an applied magnetic field is not consistent with what is observed so far in this kind of systems. Here we use single-crystal neutron diffraction to elucidate the evolution of the magnetic structure under applied magnetic field and determine the mechanism of magneto-electric coupling in different regions of the *B-T* phase diagram, which allows us to describe an unprecedented change from spin current to spin-dependent *p-d* hybridization mechanism.

A large single crystal of $(ND_4)_2[FeCl_5(D_2O)]$ of dimensions 5 x 3 x 2 mm along the crystallographic *a*-, *b*- and *c*- directions was obtained by the seeded growth technique as described in our previous work[26]. Neutron diffraction data were collected in the high resolution and low background D10 instrument at the Institut Laue-Langevin (Grenoble, France), operating in 'normal beam' geometry and equipped with a vertical-field cryomagnet, with selected applied magnetic fields ranging from 0 to 9.5 T along the *a*- and *c*- crystallographic directions. The crystal alignment with an accuracy better than 0.5 degrees was obtained by a previous orientation of the crystal using the neutron Laue diffractometer Orient Express[27]. The neutron diffraction measurements consisted initially on *Q*-scans as a function of the external field, **B**, at 2 K, for two different crystal orientations. The reversibility of the phase transitions was evaluated by collecting data increasing and decreasing magnetic field. After the exploration of the different phases as a function of the magnetic field, full data sets at selected magnetic fields were also recorded.

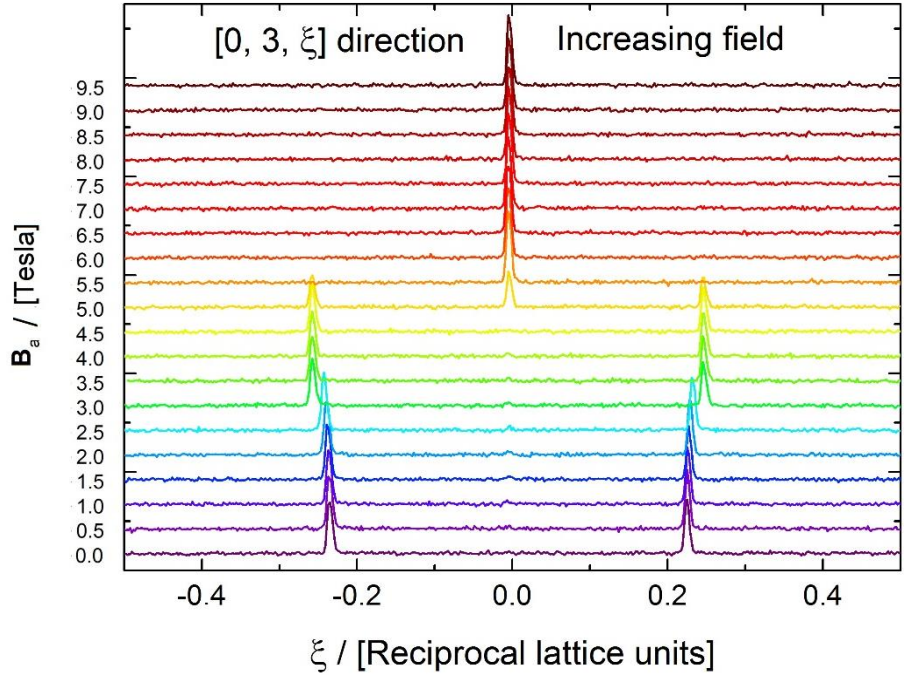

FIG. 1. *Q*-scans along the [0, 3, ξ] direction as a function of an increasing magnetic field applied along the crystallographic *a*-axis recorded at 2 K.

*Q*-scans along the [0, 3, ξ] direction as a function of the applied magnetic field (Fig. 1) give information about the complete sequence of phase transitions. At zero field, the magnetic satellites (0, 3, ± $k_z$) with $k_z$ = 0.23 correspond to the incommensurate magnetic propagation vector parallel to the *c*-axis, **k** = (0, 0, 0.23), described previously [26]. It should be noted that only primary incommensurate satellites are observed in these *Q*-scans (*Q*-scans along different directions can be consulted in Fig. S1). No sign of coexistence of incommensurate [**k** = (0, 0, ± $k_z$) with $k_z$ = 0.23] and commensurate [**k** = (0, 0, ± $k_z$) with $k_z$ = ¼] propagation vectors is observed at zero field (Figs. 1 and S1), in contrast with the preliminary results of W. Tian et al.[28] However, when the external magnetic field is applied along the crystallographic *a*-axis, the position of the magnetic satellites evolves gradually, with $k_z$ progressively increasing until it locks-in to a commensurate $k_z$ = ¼ for fields between 2.5 and 5 Tesla (Fig. 1). This first transition corresponds to the phase transition observed by macroscopic measurements between the ferroelectric phases I and II [24] (Fig. 3), characterized by a slight change in the polarization direction towards the *b*-axis, but with **P** remaining mainly directed along the *a*-axis. When the magnetic field is increased above 5 Tesla, magnetic intensity starts to appear in the otherwise very weak (0, 3, 0) nuclear reflection (see Fig. S4), indicating the prevalence of a magnetic structure with propagation vector **k** = (0, 0, 0). The intensity of this reflection becomes rapidly saturated upon further increase of the magnetic field, while the magnetic satellites (0, 3, ± $k_z$) disappear. This change corresponds, in turn, with the transition between the ferroelectric phases II and III implying an abrupt reorientation of **P** towards the *c*-axis [24] (Fig. 3). While the first transition is gradual, the first order character of the second one is clearly indicated by the hysteresis observed in the measurements with decreasing magnetic field (Fig. S2 in the supplementary material).

The series of *Q*-scans performed with the crystal oriented with the *c*-axis parallel to **B** shows qualitatively the same results - not surprisingly, given the close similarity of the *B-T* phase diagrams for both orientations of the magnetic field [24]. The resolution is however markedly poorer with **B**//c, as the experimental geometry ('normal beam') implies measurements with a departure from the horizontal scattering plane (Fig. S3 in supporting information)

The magnetic structure at 2 K under external magnetic fields of 3.5 and 6 Tesla along the *a*-axis was determined on the $(ND_4)_2[FeCl_5 \cdot D_2O]$ compound using 45 and 51 reflections, respectively. Based on the previous survey of the reciprocal space as a function of **B**, the magnetic structure in these phases is described using $\mathbf{k}_1$ = (0, 0, ¼) and/or $\mathbf{k}_2$ = (0, 0, 0) propagation vectors for 3.5 and 6 Tesla, respectively. Keeping the same nomenclature that in previous paper [26], the four Fe(III) atoms in the primitive unit-cell labelled as Fe(1), Fe(2), Fe(3) and Fe(4) correspond with crystallographic coordinates: (*x*, *y*, *z*) = (0.388, 0.249, 0.313), (1/2-*x*, 1-*y*, 1/2+*z*) = (0.119, 0.751, 0.813), (1-*x*, 1-*y*, 1-*z*) = (0.612, 0.751, 0.687) and (1/2+*x*, *y*, 1/2-*z*) = (0.881, 0.249, 0.187). The magnetic moment [$m_l(j)$] for atoms at positions Fe(j) *(j=1 to 4)* in the unit-cell *l*, can be calculated as:

$$\mathbf{m}_l(j) = \sum_{\{\mathbf{k}\}} \mathbf{S}_{\mathbf{k}j} e^{-2\pi i \mathbf{k} \mathbf{R}_l} = \mathbf{R}_{\mathbf{k}_1 j} \cos[2\pi \mathbf{k}_1 \mathbf{R}_l] + \mathbf{I}_{\mathbf{k}_1 j} \sin[2\pi \mathbf{k}_1 \mathbf{R}_l] + \mathbf{R}_{\mathbf{k}_2 j}$$

where $\mathbf{R}_l$, is the position vector of the unit-cell *l* with respect to the origin; $R_l = l_1 \mathbf{a} + l_2 \mathbf{b} + l_3 \mathbf{c}$, with $l_i$ integers, The Fourier coefficients can be written as $\mathbf{S}_{\mathbf{k}j} = 1/2(\mathbf{R}_{\mathbf{k}j} + i\mathbf{I}_{\mathbf{k}j})$ when **k** is not equivalent to –**k** and $\mathbf{S}_{\mathbf{k}j} = \mathbf{R}_{\mathbf{k}j}$ when **k** is equivalent to –**k** (like in case of **k**=1/2**H**, being **H** a reciprocal lattice vector). The last term in the previous equation has been included to take into consideration the observed field-induced ferromagnetic component corresponding with a propagation vector $\mathbf{k}_2$ = (0, 0, 0). This equation can be notably simplified in the 6 Tesla phase, where only the $\mathbf{k}_2$ = (0, 0, 0) propagation vector describes the magnetic structure. In such a case, the magnetic moment in whatever position in the crystal is identical to the magnetic moment within the unit cell:

$$\mathbf{m}_l(j) = \mathbf{R}_{\mathbf{k}_2 j}$$

In order to determine the possible magnetic structures compatible with the crystallographic space group at low temperature ($P112_1/a$) and with the mentioned propagation vectors, we have used the representational analysis as described initially by Bertaut [29]. The representational analysis for the case of $\mathbf{k}_1 = (0, 0, ¼)$, involves two one-dimensional irreducible representations (*irreps*), $\Gamma_1$ and $\Gamma_2$, of the *little group* $G_{\mathbf{k}1}=(P112_1)$. In this symmetry group, the Fe(III) site is split into two different orbits, as occurs at zero field [26]. The two orbits can be merged into the original orbit when we consider the *extended little group* $G_{\mathbf{k}1,-\mathbf{k}1}=(P112_1/a)$, coincident with the space group. The magnetic representation can be decomposed in *irreps* of $G_{\mathbf{k}1}$, $\Gamma = 3\Gamma_1 + 3\Gamma_2$, so that three sets of basis vectors for each *irrep* can be considered in the most general (lower symmetry) case. As in the zero field phase, none of these *irreps* alone is able to generate the experimentally observed macroscopic electric polarization [24]. In order to allow the occurrence of electric polarization and fit properly the neutron diffraction data, a combination of magnetic models belonging to the $\Gamma_1$ and $\Gamma_2$ irreducible representations is necessary. Moreover, in order to include the effect of the magnetic field, a weak ferromagnetic component along the *a*-axis (direction of application of the external magnetic field) has been included in the model. The value of the field-induced ferromagnetic moment was derived from the magnetization curves as function of the external field [24], with an estimated value of 0.4 $\mu_B$ along the *a*-axis.

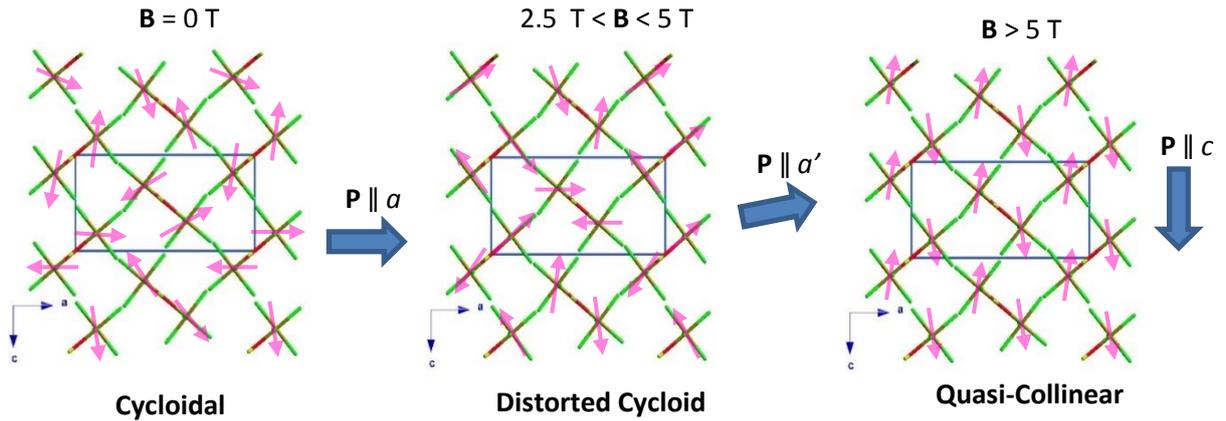

FIG. 2. Magnetic structures at 2 K and at different magnetic fields applied along the crystallographic *a*-axis: at zero field (left), at 3.5 (middle), and at 6 Tesla (right).

The refined magnetic model can be seen as a distorted cycloidal magnetic structure due to the superposition of a cycloidal structure plus a ferromagnetic component. The magnetic model shows two remarkable differences with respect to the previously obtained at zero field [26]. The first one is related with the orientation of the magnetic moments, which are significantly out of the *ac*-plane. The tilting angle ranges from 17 to 30 deg., values which are notably larger than the previously reported of 4 deg. determined for the cycloidal arrangement at zero field. The second difference is related with the effect of the ferromagnetic component along the *a*-axis in the cycloid structure. This ferromagnetic component produces a reorientation of the magnetic moments, which is also compatible with the occurrence of a reorientation on the electric polarization [30]. A schematic representation of this magnetic structure can be seen in Fig. 2. It deserves to be noted that the ferromagnetic signal is very weak and therefore the influence in the pure cycloidal model is very subtle.

The representational analysis for the case of **k** = (0, 0, 0) gives four one dimensional irreducible representations. The magnetic representation $\Gamma$ for the general Wyckoff position 4*e* can be decomposed as a direct sum of these irreducible representations. The four possible irreducible representations for the 4*e* site ($\Gamma_1$-$\Gamma_4$) correspond with the magnetic space groups $P112_1/a$, for $\Gamma_1$, $P112_1/a'$, for $\Gamma_2$, $P112_1'/a'$, for $\Gamma_3$ and $P112_1'/a$, for $\Gamma_4$. The only magnetic space groups compatible with the lineal magneto- electric behavior observed in the electric polarization measurements [24] are those derived from $\Gamma_2$ and $\Gamma_4$. However, none of these solutions reproduce the experimental data. For $\Gamma_2$ the data refinement leads to poor agreement factors (above 20 % $R_F^2$), with large errors in the refined magnetic moments, which are meaningless. The refinement using $\Gamma_4$ gives a reasonable goodness of fit, but this model alone is not compatible with the non-negligible field-induced ferromagnetic component along the *a*-axis, which was observed by macroscopic magnetometry measurements [24]. The magnetic structure derived from $\Gamma_4$ implies a strictly antiferromagnetic coupling of the magnetic moments along this direction and consequently this model is not suitable to account for the experimental results. A combination of magnetic modes allowing a ferromagnetic component along the *a*-axis is therefore necessary to reproduce the experimental data. The only combination that fits properly the data is the admixture of $\Gamma_4$ and $\Gamma_3$ (the last one being the only *irrep* giving a ferromagnetic component in the *a*-axis), resulting in the $P112_1'$ magnetic space group. The magnetic arrangement can be seen as a canted structure where the magnetic moments are mainly contained in the *bc*-plane and antiferromagnetically coupled (Fig. 2), with the main direction pointing along the *c*-axis. The application of the external magnetic field produces a tilt of the magnetic moments towards the *a*-axis, giving rise to a net ferromagnetic signal. The value of this ferromagnetic component can be estimated from magnetometry measurements to be of *ca.* 0.9 $\mu_B$ (per iron site). Due to the small value of this component, and in order to avoid over-parametrization, this value was fixed in the neutron data refinement. Under these conditions, the refined Fe(III) magnetic moment is 4.6 (3) $\mu_B$, remarkably larger than the value obtained at zero field in the cycloid magnetic structure [26]. The increase of magnetic moment in this phase should be related with the decrease of magnetic frustration from the cycloidal magnetic structure to the canted antiferromagnetic structure. The obtained value of the magnetic moment is in good agreement with those previously reported for other members of this series [31].

The proposed magnetic structures fully explain the observations in the electric polarization. At 2 K and 3.5 Tesla, the electric polarization is explained by the spin current mechanism, similarly to the situation at zero field. As mentioned before, the deformation of the cycloids due to the field-induced ferromagnetic component is also compatible with the observed magneto-electricity [30]. The observation of **P** preferentially directed along the $a$-axis [24] is a consequence of the resulting fan-like structure, with moments rotating mainly in the $ac$-plane. The tilt of the fan rotation plane towards $b$-axis is consistent with the weak electric polarization component along the $b$-axis, and the variation of the tilting angle with respect to the value at zero field is the responsible of the reorientation of the electric polarization observed in the macroscopic measurements [24].

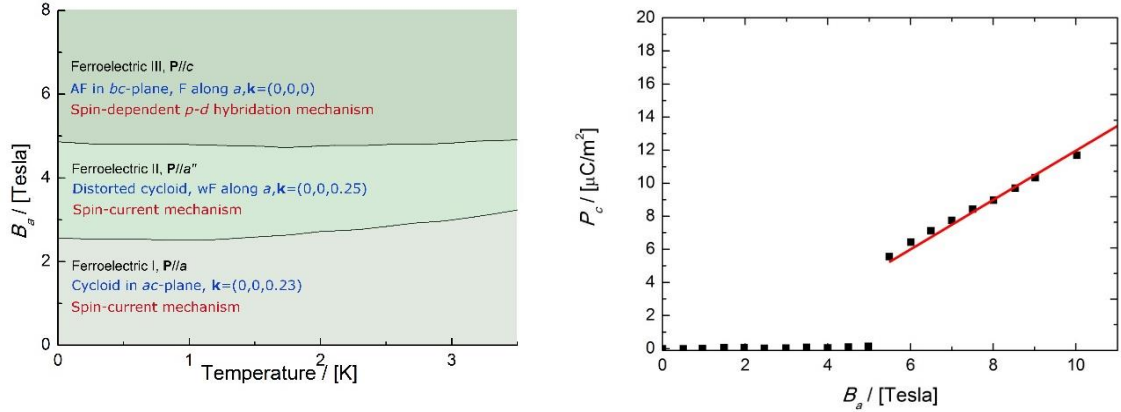

FIG. 3. (Left) $B$-$T$ phase diagram for magnetic field applied along the crystallographic $a$-axis, with a summary of the different ferroelectric phases (black), magnetic structures (blue) and magneto-electric coupling mechanisms (red). The phase boundaries have been taken from ref. 24. (Right) Dependence of the component of electric polarization along the crystallographic $c$-axis for a magnetic field applied along the $a$-axis. The symbols correspond to the measurements reported by Ackermann et al. (data taken from ref. 24), while the continuous line represents a calculation of the polarization predicted by the p-d hybridization model, assuming a linear dependence with **B** of the component of the magnetic moment along the $a$-axis, together with a constant value for the component in the $bc$-plane.

In contrast, at 2 K and 6 Tesla, the observed polarization can be explained by the spin-dependent p-d hybridization model. Given the crystal structure of $(ND_4)_2[FeCl_5·D_2O]$ and the proposed magnetic arrangement, one can calculate the product $(\mathbf{S}_i \cdot \mathbf{e}_{il})^2 \mathbf{e}_{il}$, which would be proportional to the local polarization along the bond directions $\mathbf{e}_{il}$. Summing for all the bond directions ($l$ = 1 to 6) and for the four Fe(III) sites ($i$ = 1 to 4) it is found that the proposed magnetic structure produces a net electric polarization along the $c$-axis, in agreement with the macroscopic observations. Likewise, it can be also seen that for other possible magnetic arrangements ($\Gamma_1$ and $\Gamma_4$ and combinations of them), no net polarization is produced. Furthermore, the evolution with **B** of the polarization component along $c$-direction in this phase can be qualitatively reproduced (Fig. 3) if we assume a linear dependence with **B** of the component of the magnetic moment along the $a$-axis, together with a constant value for the component in the $bc$-plane. The latter assumption is supported based on the non-variation of the intensity of the (0, 3, 0) reflection, which is mainly sensitive to the magnetic moments variation on the $c$-axis. A calculation of the intensity of this reflection as function of the tilting angle of the magnetic moments into the $ac$-plane can be consulted in Fig. S4. The deviation of the magnetic moments ca. 10 deg. from the quasi-collinear structure (as obtained at 2K and 6 T) produces a variation of ca. 3% in the intensity of the (0, 3, 0) reflection, which is in the limit of precision of these measurements.

Therefore, our results describe an unprecedented example of change in the mechanism of induced electric polarization, upon application of magnetic field, from spin current to spin-dependent $p$-$d$ hybridization mechanism.

In summary, we have followed the evolution of the magnetic structures of $(ND_4)_2[FeCl_5 \cdot D_2O]$ compound using single crystal neutron diffraction under external magnetic field. The obtained results allowed us to explain the different ferroelectric phases based on the changes in the magnetic structure. The observed electric polarization has been explained using two different magneto-electric coupling mechanisms: the spin-current mechanism for external magnetic field below 5 T, and the spin dependent $p$-$d$ hybridization mechanism for magnetic field above this value, being this compound the first example reported presenting this sequence of magneto-electric coupling mechanisms.

Partial funding for this work is provided by the Ministerio Español de Ciencia e Innovación through projects MAT2010-16981, MAT2011-27233-C02-02. JARV acknowledges CSIC for a JAEdoc contract. We are grateful to Institut Laue-Langevin for the neutron beam-time allocated through project "5-41-770".

# Magnetic-Field-Induced Change of Magneto-Electric Coupling in the Molecular Multiferroic $(ND_4)_2[FeCl_5(D_2O)]$

**J. Alberto Rodríguez-Velamazán,**[1,2] **Oscar Fabelo,**[1] **Javier Campo,**[2] **Angel Millan,**[2] **Juan Rodríguez-Carvajal,**[1] **and. Laurent C. Chapon**[1]

[1] *Institut Laue-Langevin, 71 Avenue des Martyrs, CS 20156, 38042 Grenoble Cedex 9, France.*
[2] *Instituto de Ciencia de Materiales de Aragón, CSIC-Universidad de Zaragoza, C/ Pedro Cerbuna 12, E-50009, Zaragoza, Spain.*

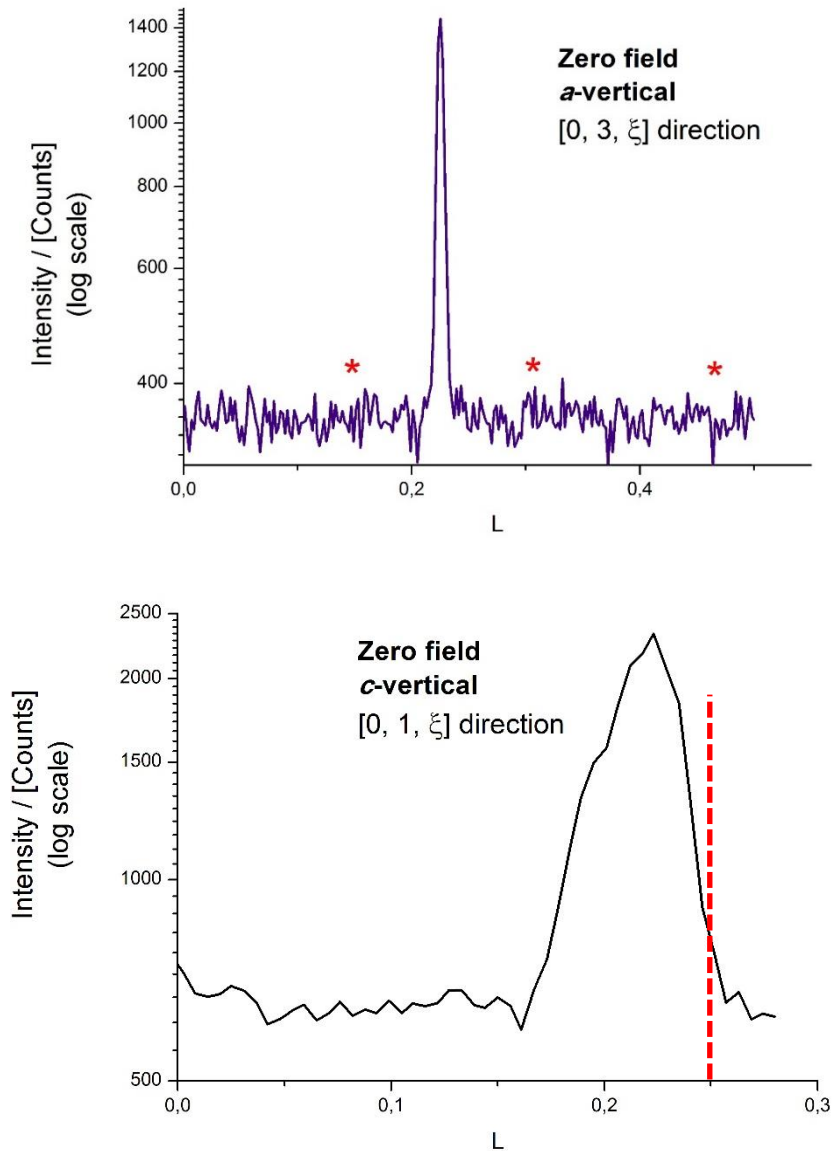

FIG. S1. (top) Q-scan along the [0, 3, ξ] direction at 2 K and zero magnetic field. The red asterisks denote the positions of the possible incommensurate satellites, corresponding to -5k, -3k and 2k, from left to right, respectively. (bottom) Q-scan along the [0, 1, ξ] direction at 2 K and zero magnetic field. The red dashed line represents the calculated position for the first order (primary) commensurate structure with **k** = (0, 0, ¼). Logarithmic scales have been used in both figures to enlarge possible low intensity signals.

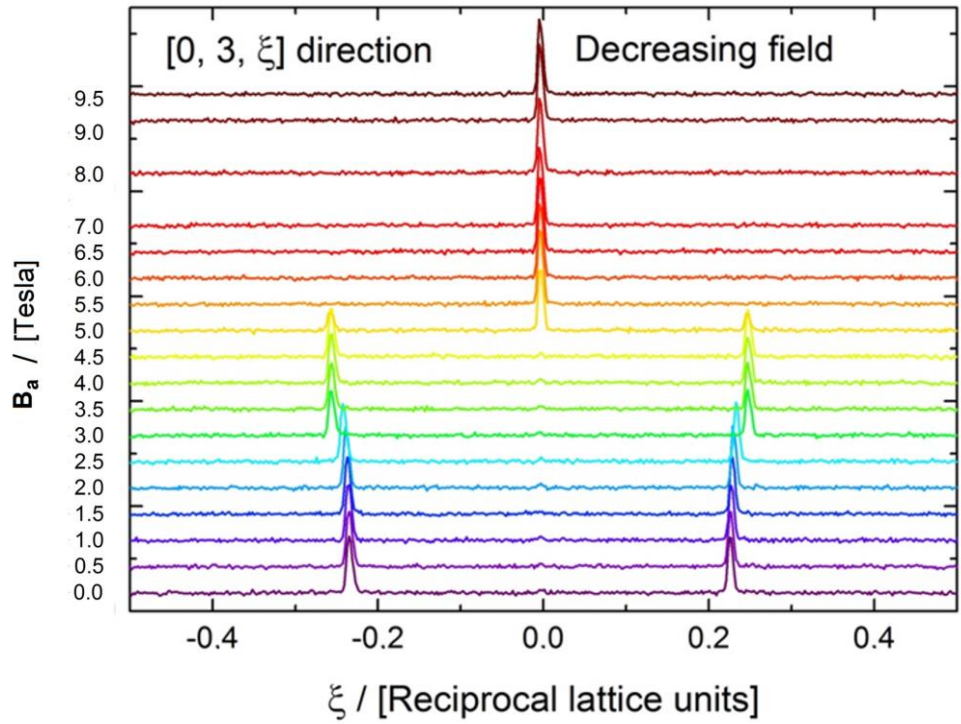

FIG. S2. *Q*-scans along the [0, 3, ξ] direction as a function of a decreasing magnetic field applied along the crystallographic *a*- axis recorded at 2 K.

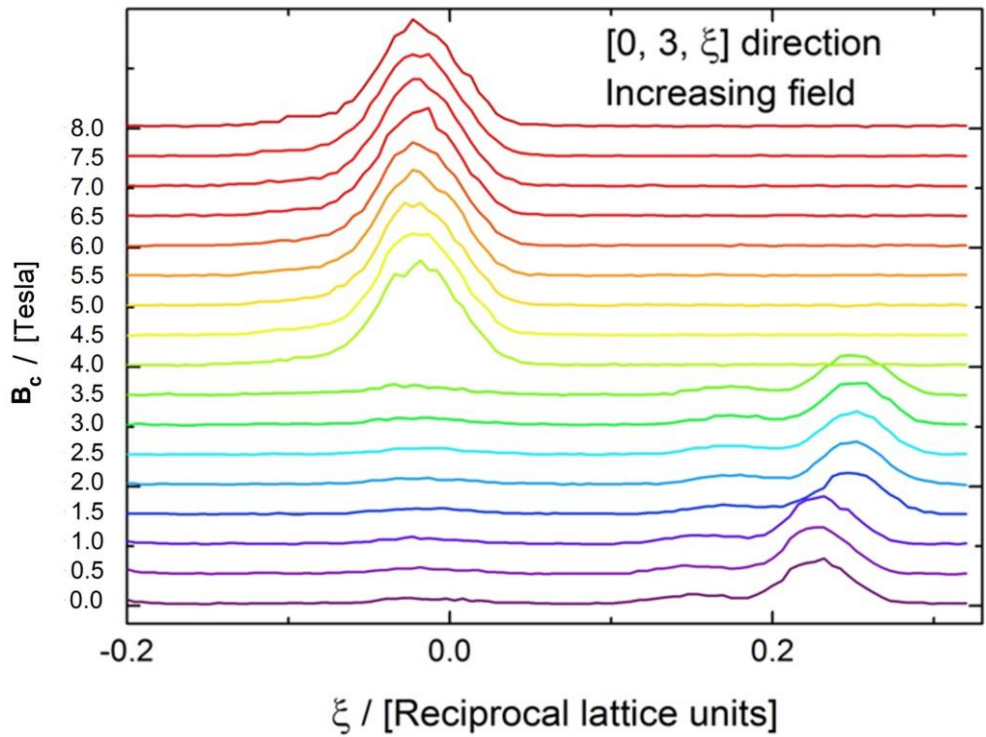

FIG. S3. Q-scans along the [0, 3, ξ] direction as a function of an increasing magnetic field applied along the crystallographic *c*- axis recorded at 2 K.

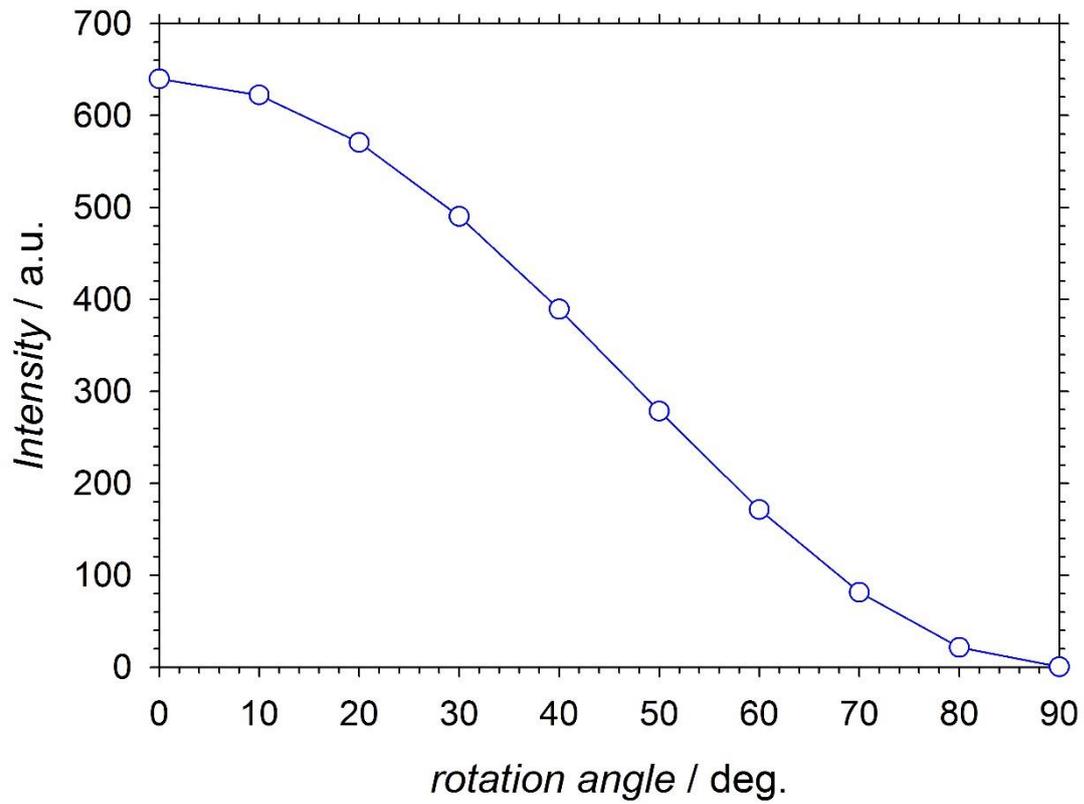

FIG. S4. Calculation of the intensity of the (0, 3, 0) reflection as function of the rotation angle along the *b*-axis, being 0 deg the *c*-axis and 90 deg. the *a*-axis. The value of the magnetic moments have been fixed to the value obtained from the data refinement at 6 Tesla and 2K.